\documentclass[conference,letterpaper]{IEEEtran}
\IEEEoverridecommandlockouts
\usepackage{cite}
\usepackage{amsmath,amssymb,amsfonts}
\usepackage{algorithmic}
\usepackage{graphicx}
\usepackage{textcomp}
\usepackage{xcolor}
\usepackage{float}
\usepackage{multirow}
\usepackage{url}

\def\BibTeX{{\rm B\kern-.05em{\sc i\kern-.025em b}\kern-.08em
    T\kern-.1667em\lower.7ex\hbox{E}\kern-.125emX}}

\usepackage{longtable}
\usepackage{comment}
    
\begin{document}

\title{Self-Supervised Learning for Android Malware Detection on a Time-Stamped Dataset\\
{\footnotesize }
\thanks{The authors would like to thank Mastercard
Canada for supporting this research.}
}

\author{\IEEEauthorblockN{Annan Fu}
\IEEEauthorblockA{\textit{Khoury College of Computer Sciences} \\
\textit{Northeastern University}\\
Vancouver, Canada \\
fu.anna@northeastern.edu}
\and
\IEEEauthorblockN{Hao Pei}
\IEEEauthorblockA{\textit{Khoury College of Computer Sciences} \\
\textit{Northeastern University}\\
Vancouver, Canada \\
pei.h@northeastern.edu }
\and
\IEEEauthorblockN{Maryam Tanha}
\IEEEauthorblockA{\textit{Khoury College of Computer Sciences} \\
\textit{Northeastern University}\\
Vancouver, Canada \\
m.tanha@northeastern.edu}
}

\maketitle

\begingroup
\renewcommand\thefootnote{}
\footnotetext{
© 2026 IEEE. Personal use of this material is permitted. Permission from IEEE must be obtained for all other uses, 
in any current or future media, including reprinting/republishing this material for advertising or promotional purposes, 
creating new collective works, for resale or redistribution to servers or lists, or reuse of any copyrighted component 
of this work in other works.

This is the author’s accepted manuscript. The final version of record will be available in IEEE Xplore.
}
\endgroup

\begin{abstract}
Android malware detectors built with machine learning often suffer from temporal bias: models are trained and evaluated without respecting apps’ actual release times, inflating accuracy and weakening real-world robustness. We address this by constructing a time-stamped dataset of benign and malicious Android apps and introducing a timestamp-verification procedure to ensure temporal accuracy. We then propose a detection framework that uses Bootstrap Your Own Latent (BYOL) for self-supervised pre-training to learn obfuscation-resilient representations, followed by supervised classification. Under time-aware evaluation, the method attains 98\% accuracy and 89\% F1. We further characterize malware behavior by analyzing true positives and false negatives using VirusTotal and the MITRE ATT\&CK framework. To support reproducibility and further innovation, we release our dataset and source code.
\end{abstract}

\begin{IEEEkeywords}
Android, malware, self-supervised learning, temporal bias
\end{IEEEkeywords}

\section{Introduction}
Machine learning (ML) has been widely used for Android malware detection; however, such models face critical challenges. First, concept drift, i.e., distribution shifts over time from evolving malware families, benign apps, or platform features—can cause models trained on historical data to degrade, a phenomenon known as model aging. The second issue is temporal bias (also called temporal inconsistency) that stems from not considering chronological order for train/test splits during evaluation. This can lead to over-optimistic results that fail to reflect real-world scenarios~\cite{guerra2024machine}\cite{guerra2021kronodroid}\cite{pendlebury2019tesseract}. To properly address both concept drift and temporal bias in Android malware detection, we should have a reliable timestamp for each app~\cite{tanha2024revisiting}. However, finding a guaranteed ground-truth timestamp is challenging. This is due to the fact that such metadata (e.g., compilation time) can be easily modified, either intentionally by attackers  or because of development mistakes or even may not be available for all apps (e.g., \textit{uploadDate} from AndroZoo~\cite{Allix:2016:ACM:2901739.2903508}).

Furthermore, Android malware data has a few practical problems: the labels are not always reliable (heuristic VirusTotal~\cite{VirusTotal} votes, malware shows different behavior at run-time vs. static analysis), many benign and malicious apps look similar in their API calls, permissions, or opcodes (so, it is hard to pick clear “opposites” for contrastive learning), and the data shifts over time because malware evolves and uses obfuscation. As a result, fully supervised models can overfit to inaccurate labels, and contrastive methods assume clear semantic separation between classes, which breaks down when there is high feature overlap or when ``hard negative samples" (similar-but-different samples) dominate.


Considering the aforementioned issues, in this paper, we first create a time-stamped dataset of Android goodware and malware as well as proposing a timestamp verification method using the API call introduction time. Moreover, we adopt Bootstrap Your Own Latent (BYOL)~\cite{grill2020bootstrap} as a self-supervised pre-training framework to enhance feature representations before supervised classification. BYOL leverages all data, regardless of label quality, to learn generalized and behaviorally meaningful representations. Unlike contrastive methods, BYOL does not require explicit negative pairs, avoiding instability caused by overlapping API calls or permission patterns between apps. Overall, BYOL learns obfuscation-resilient and behaviorally consistent embeddings from unlabeled features that improve robustness to unseen malware and mitigate concept drift.

This paper makes the following key contributions:
\begin{itemize}
    \item \textbf{Creating a time-stamped dataset}: We have created a time-stamped dataset of Android benign applications and malware samples. This time-based dataset helps tackle the temporal bias in ML-based Android malware detectors. Additionally, since app time-related metadata may not reliably reflect its release date, we introduce a timestamp verification method to improve temporal accuracy.
    \item \textbf{Using self-supervised learning:} To the best of our knowledge, we are the first to employ the BYOL self-supervised learning framework to enhance the ability of Android malware detectors to identify new or obfuscated malicious apps. 
    \item \textbf{Open Science:} We have made our dataset and code publicly available on GitHub repository\footnote{https://github.com/maryam-tanha/Self-supervised-Learning-for-Android-Malware-Detection} to facilitate reproducibility and further research in Android malware detection. 
   
\end{itemize}
 
 The rest of the paper is organized as follows. Section~\ref{sec:related word} provides a review of existing studies on addressing temporal bias as well as the works on self-supervised learning for Android malware detection. Our methodology, including the creation of our time-stamped dataset, timestamp verification process, and malware detection using BYOL and logistic regression (LR) is discussed in Section~\ref{sec:methodology}. The performance evaluation of our proposed system and analysis of the results are included in Section~\ref{sec:eval}. Finally, Section~\ref{sec:conclusion} concludes the paper.

\section{Related Work}
\label{sec:related word}
We reviewed existing studies on Android malware detection with respect to the following aspects. 
\begin{enumerate}
\item \textbf{Concept drift and temporal bias}:  
TESSERACT~\cite{pendlebury2019tesseract} is an open-source evaluation framework designed to eliminate spatial and temporal experimental bias in ML-based Android malware detection. DroidEvolver~\cite{xu2019droidevolver} mitigates concept drift without full retraining or true labels by using lightweight online updates with evolving feature set and pseudo labels. It flags drifting apps via a Juvenilization Indicator. APIGraph~\cite{zhang2020enhancing} captures semantic similarity between APIs and improves state-of-the-art classifiers so that they can still detect evolved malware.
KronoDroid~\cite{guerra2021kronodroid} is a hybrid-featured Android dataset (a combination of static and dynamic features) and provides timestamps for the apps to facilitate the detection and characterization of the evolution of Android malware (concept drift). Explainable AI (XAI) methods are employed in~\cite{liu2022explainable} to investigate the over-optimistic performance of Android malware classifiers. It shows that such high performance is due to the fact that ML models learn to distinguish apps based on temporal differences rather than actual malicious behaviors. Debiasing algorithms are proposed in~\cite{miranda2022debiasing} to be applied to data sets to ensure that they are representative of the population at the time of the experiment, mitigating both concept drift and temporal bias. 
A practical way to handle concept drift is classification with rejection, i.e., quarantine samples likely to be misclassified for expert review. This idea is used in~\cite{barbero2022transcending} by introducing a framework called TRANSCENDENT (built on Transcend~\cite{jordaney2017}). The authors show that their proposed framework outperforms state-of-the-art baselines and generalizes across malware classifiers. To properly address both temporal bias and concept drift, we should assign a timestamp to each app that reflects its release time to the market. In the existing studies, two main approaches are utilized as follows:
\begin{itemize}
\item  \textbf{Internal Timestamps (extracted from the application package(APK) File):} Internal timestamps are metadata extracted directly from the application package or the files it contains. These are often manipulable by attackers or developers~\cite{li2018moonlightbox}\cite{guerra2021kronodroid}. The first timestamp in this category is \textit{compilation date (or DEX date)} that approximates the assembly time by checking the last-modified time of the main DEX file. However, it is now considered unusable because recent apps frequently provide an invalid value, such as 1980~\cite{tanha2024revisiting}. Other timestamps are \textit{Last Modification} and \textit{Earliest Modification}. The former is the latest modification timestamp found in any of the inner files that compose the application while the latter is the earliest modification timestamp found in any of the inner files of the application~\cite{guerra2021kronodroid}. 

\item \textbf{External Timestamps (extracted from VirusTotal/Online Sources):} External timestamps are derived from sources outside the control of the application developers/attackers, making them generally more robust against tampering. The most common one is \textit{First Seen VT (VirusTotal Submission)}~\cite{guerra2024machine}. However, this timestamp does not necessarily reflect the app release time to the market. One of the recent studies~\cite{tanha2024revisiting} leveraged the Google Play Store \textit{uploadDate} metadata from AndroZoo~\cite{Allix:2016:ACM:2901739.2903508} as a precise indicator of an app's release time. By relying on this timestamp, the authors created a temporally-consistent dataset. However, \textit{uploadDate} is not available for all the apps from AndroZoo (only available for the apps from Google Play market).
\end{itemize}
While we used the second approach to assign timestamps to the apps in our dataset, we have proposed a verification algorithm for the assigned timestamps. 

\item \textbf{Self-supervised learning}:
The use of self-supervised learning has shown promises in improving the performance of Android malware detectors\cite{chen2023continuous,wu2022contrastive,seneviratne2022self}. One of the key studies is~\cite{chen2023continuous} which addresses concept drift by combining active learning with a hierarchical contrastive encoder and a new pseudo-loss uncertainty score for sample selection. This approach maintains more stable accuracy over several years and reduces analyst labeling efforts. IFDroid~\cite{wu2022contrastive}, is a framework that enhances Android malware family classification using contrastive learning and function call graph (FCG) analysis. The FCGs of the apps are converted into images. Then, contrasive learning is performed on these images. This framework achieves robust, obfuscation-resilient, and interpretable Android malware classification. Complementary to this line of work, our approach leverages BYOL, a non-contrastive self-supervised method that learns without requiring negative samples, to extract obfuscation-resilient representations from unlabeled data. SHERLOCK~\cite{seneviratne2022self} is a self-supervised deep learning framework for malware detection built upon the Vision Transformer architecture. It adopts an image-based binary representation of executables to automatically learn distinctive features that distinguish malicious programs from benign ones. The performance evaluation of SHERLOCK shows its effectiveness in binary and multi-class Android malware classification. 

\end{enumerate}
\section{Methodology}
\label{sec:methodology}
\begin{figure*}[t]
  \centering
  \includegraphics[width=\linewidth]{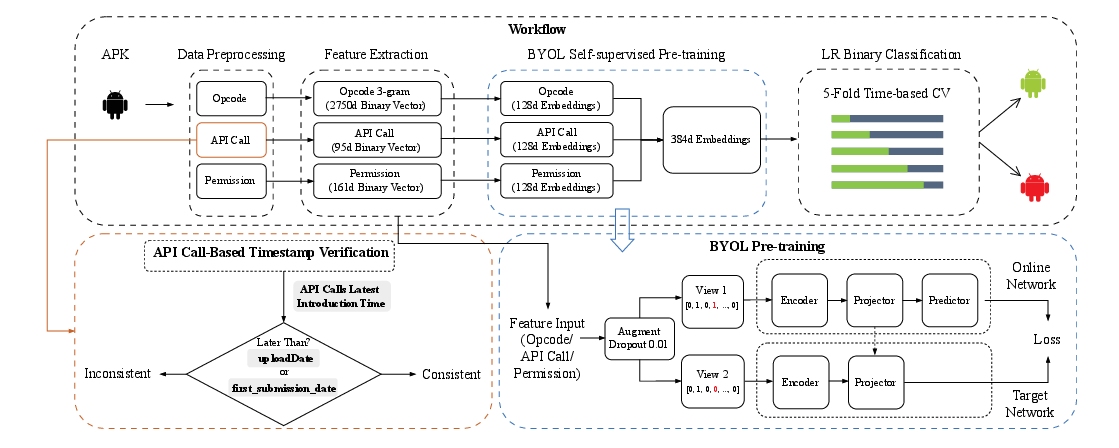}
  \caption{Overview of Our Methodology}
  \label{fig:workflow}
\end{figure*}

\normalsize
Our methodology follows a systematic workflow as shown in Figure~\ref{fig:workflow}.


\subsection{Time-Based Dataset Construction and Organization}
Our dataset construction addresses temporal inconsistency by ensuring chronological integrity between training and test data. We created a dataset of 40,000 Android applications spanning the year 2021 to 2024, maintaining a realistic 9:1 ratio of benign to malware samples to reflect real-world distributions.

\subsubsection{Temporal Splitting Strategy} To mitigate temporal bias, our methodology implements a chronological splitting strategy. The splitting approach ensures that all test samples represent future time periods relative to training samples.

The test set comprises 4,000 samples (400 malware and 3,600 benign) exclusively from 2024, representing the most recent applications in our dataset. The training set contains 36,000 (3,600 malware and 32,400 benign) samples from applications time-stamped between 2021 and 2024. This temporal arrangement guarantees that all training data precedes test data chronologically, eliminating the possibility of training on ``future'' information.

\subsubsection{Dataset Collection} We collected benign APKs exclusively from AndroZoo using \textit{vt\_detection} = 0 and \textit{market} from GooglePlay as filters. Timestamps were obtained from AndroZoo's \textit{uploadDate} metadata as a reliable timestamp for chronological ordering~\cite{tanha2024revisiting}. We initially collected malware from AndroZoo using $\textit{vt\_detection} \ge $15 as the primary identification criterion\cite{chen2023continuous}. For samples lacking \textit{uploadDate} information, we utilized VirusTotal's \textit{first\_submission\_date} metadata as the timestamp. Due to limited recent malware samples, we supplemented our dataset with malware samples from MalwareBazaar\cite{malware-bazaar}, utilizing VirusTotal's \textit{first\_submission\_date} metadata as timestamps. To ensure consistency, all malware samples selected from MalwareBazaar were required to have a VirusTotal detection count of at least 15. Table~\ref{table:dataset_distribution} presents the detailed distribution of our time-stamped dataset.

\begin{table}[ht]
\caption{Dataset Distribution}
\centering
\scalebox{0.9}{
\begin{tabular}{ccccc}
\hline
Year & Benign & Malware & Malware & Total \\
& (uploadDate) & (uploadDate) & (first\_submission\_date) & \\
\hline
2024 & 9,000 & 17 & 383 & 9,400 \\
2023 & 9,000 & 97 & 114 & 9,211 \\
2022 & 9,000 & 248 & 767 & 10,015 \\
2021 & 9,000 & 426 & 1,948 & 11,374 \\
\hline
Total & 36,000 & 788 & 3,212 & 40,000 \\
\hline
\end{tabular}
}
\label{table:dataset_distribution}
\end{table}

\subsubsection{Timestamp Verification} Although the \textit{uploadDate} provided by Google (and available via AndroZoo) serves as a good indicator of an app’s market release time, it is not always available. Similarly, the \textit{first\_submission\_time} reported by VirusTotal does not necessarily reflect the actual release date of an app. Therefore, we developed a timestamp verification process to address this limitation. Building upon the methodology of~\cite{li2018moonlightbox}, we developed a systematic API call-based verification method applied to all 40,000 samples. We first extracted API calls used by each app using static analysis (Androguard tool~\cite{AndroG}). We then estimated each app’s market release time (a lower bound) by taking the most recent introduction timestamp among its API calls and compared it with \textit{uploadDate} from AndroZoo and \textit{first\_submission\_date} from VirusTotal. The detailed procedure is described below.

 Our primary data source for API call timestamp analysis consists of API call first introduction timestamps systematically extracted from the AOSP Git Repository across Android versions 1-15~\cite{andro-plat}. Then, for each API class–method signature, we located all repository versions in which that API appears. Subsequently, we selected the earliest timestamp as the API call's first introduction date and added it to a table recording the first appearance of each API call. This table is used to identify the API call with the most recent introduction timestamp for an app as a lower bound for its market release date. To do so, direct matching by exact class-method signatures may fail due to Java inheritance mechanisms, where applications can invoke APIs through parent classes or interfaces rather than the specific implementing class. To address this challenge, we developed a multi-level matching strategy supported by creating additional tables based on~\cite{andro-plat, andro-dev}. The key tables are \textit{inheritance hierarchy mapping} for class relationships, and \textit{API level to package-class-method mapping} with corresponding Android version release dates for complementary API call timestamp matching. If direct API call matching fails, inheritance hierarchy-based matching is used. If the latter step fails as well, we adopt API level mapping and the corresponding Android version release dates. Using this process, we identified the introduction timestamp for all of the standard API calls extracted from all the apps in our dataset. It is important to note that our analysis focuses exclusively on standard APIs documented in the Android Developer reference. The excluded APIs represent only 0.4\% of all API calls extracted from the apps in our dataset. 

We analyzed temporal consistency across 40,000 samples by comparing the app’s latest API call timestamp as the temporal lower bound with AndroZoo’s \textit{uploadDate} or VirusTotal’s \textit{first\_submission\_date} when  \textit{uploadDate} is not available. Temporal discrepancy occurs when the app's latest API timestamp is later than the AndroZoo \textit{uploadDate} or VirusTotal \textit{first\_submission\_date} (we compared the year only). Our analysis revealed that the temporal discrepancy rate was at most 0.02\% across all years, which confirmed both \textit{uploadDate} and \textit{first\_submission\_date} as reliable timestamps to estimate the app market release date for time-based machine learning experiments. 

\subsection{Feature Extraction}
Static analysis involves analyzing the source code of Android apps without running them. Static features are extracted from the APK files, specially the \textit{AndroidManifest.xml} and \textit{Dalvik Executable (.dex)} files. The most commonly used static features include permissions, API calls, and opcode sequences.  

\subsubsection{Opcode n-gram features} Opcodes have proven particularly valuable, as they capture low-level execution patterns that can reveal malicious behaviors and remain less dependent on high-level obfuscation. Opcodes are often processed using n-gram models (e.g., FAMCF~\cite{zhou2024famcf} and FAMD~\cite{bai2020famd}). Further analysis in~\cite{kang2016n} showed that smaller N values (3 or 4) yield the best malware classification performance. Moreover, some studies such as DeepDetect~\cite{kumar2021deepdetect} and DeepRefiner~\cite{xu2018deeprefiner} use a simplified or reduced instruction set to enhance efficiency. We employed a 17-category symbolic mapping system that extends DeepRefiner~\cite{xu2018deeprefiner}'s 15-category classification with 2 additional categories (\textit{invoke-polymorphic/custom} and \textit{const-method}) based on the latest Android version opcodes{\cite{andro-opcode}}. Android APKs were decompiled using Apktool~\cite{apk-tool} and baksmali\cite{baksmali} to extract Dalvik opcodes, which were mapped to the symbolic categories. We represented each APK as a concatenated app-level symbolic sequence and generated a binary one-hot vector of 3-gram features capturing the presence or absence of each pattern. After excluding zero-occurrence patterns from the training set, this resulted in 2,750-dimensional binary feature vectors for each application.

\subsubsection{API calls and permissions}We extracted both API calls and permissions from Android apps using Androguard. Our initial feature set included the same permissions and API calls used in~\cite{tanha2024revisiting}. We augmented the feature set with 103 warning-level permissions. These permissions were extracted from VirusTotal reports for our malware samples, retaining only standard Android framework permissions and excluding app-defined ones. API calls and permissions were encoded as 95-dimensional and 161-dimensional binary vectors, respectively, indicating the presence or absence of each feature.

\subsection{Self Supervised Learning and Binary Classification}
 AS we mentioned in Section~\ref{sec:related word}, self-supervised learning methods have shown promises for Android malware detection.
\subsubsection{Overview of the BYOL architecture and pre-training mechanism} We employed BYOL~\cite{grill2020bootstrap} as our self-supervised representation learning framework for Android app features. BYOL consists of two neural networks: an online network comprising encoder, projector, and predictor multilayer perceptrons, and a target network containing only encoder and projector components. The encoder uses a three-layer MLP architecture with batch normalization and ReLU activations between layers, while both the projector and predictor employ two-layer MLPs with batch normalization on the hidden layer. The target network parameters are updated via exponential moving average of the online network parameters, providing a slowly moving teacher signal that enables learning without requiring negative samples or contrastive pairs. During training, two augmented views are created from each input sample, with the online network trained to predict the target network's projections using cosine similarity loss between normalized prediction and target vectors. For our binary Android features (permissions/API calls/opcode 3-gram sequences), we created augmented views using feature dropout, randomly masking present features while preserving semantic validity by avoiding the introduction of non-existing features. Given the dimensional imbalance between feature types (2,750-dimensional opcode 3-gram vectors, 95-dimensional API vectors, 161-dimensional permission vectors), we trained separate BYOL models for each feature type. This prevents high-dimensional features from dominating the representation space. Each modality-specific encoder generates 128-dimensional embeddings, which are concatenated into a 384-dimensional representation, ensuring equal contribution from each feature type.

\subsubsection{Design rationale for adopting BYOL}
Although our dataset includes a substantial number of labeled benign and malicious Android apps, we adopt BYOL as a self-supervised pre-training framework to enhance the quality of feature representations prior to supervised classification. Our design choice is motivated by two key considerations. First, in Android malware analysis, labels can be noisy or inconsistent due to changing app behavior and reliance on heuristic or antivirus consensus. BYOL leverages the entire dataset, regardless of label accuracy, to learn generalized and behaviorally meaningful representations that capture latent semantic relationships among applications. Second, unlike contrastive methods, BYOL removes the need for explicit negative pairs. In malware datasets, negative samples are hard to define because different applications often share similar API calls or permissions. Without contrastive loss, BYOL avoids this issue and produces more stable embeddings. In summary, BYOL learns obfuscation-resilient and behaviorally consistent embeddings from unlabeled permission, API calls, and opcode features. This approach combines the generalization power of self-supervision with the precision of supervised learning, improving robustness to unseen malware and mitigating concept drift from evolving Android behaviors.

\subsubsection{Binary Classification with Time-based Cross-Validation}
We selected LR model as the binary classifier for malware detection due to its ability to efficiently process continuous, dense embeddings produced by BYOL. It offers linear interpretability through its coefficients, and leverages L2 regularization to mitigate overfitting. We applied inverse-frequency class weighting to address the class imbalance in our dataset. Moreover, we applied a tailored five-fold time-based cross-validation on the training set for hyperparameter tuning.

\section{Performance Evaluation and Analysis of Results}
\label{sec:eval}
We conducted our experiments on a MacBook Air with Apple M2 processor and 16 GB memory. To evaluate the effectiveness of our proposed system, the LR classifier was trained on the BYOL embeddings generated for different feature sets using five-fold cross-validation procedure with hyperparameter tuning. For the full set of BYOL hyperparameter settings, we refer the interested reader to our public GitHub repository. Table~\ref{table:performance_comparison} presents the test-set results for the LR classifier using different feature sets, alongside the prediction performance of a Random Forest (RF) classifier trained on the full feature set. 
The linear nature of LR is well-suited to the structured representation space created by BYOL pre-training. As shown in the table, when we use all of the extracted static features, i.e., permissions, API calls and opcode sequences, the best performance for LR is achieved. When trained on all features, the RF classifier attains higher precision (97\% vs. 89\%), indicating a more conservative detector with fewer false positives but more missed malware samples. However, in malware detection, we prioritize higher recall to minimize false negatives.


\begin{table}[ht]
\caption{Classification Performance}
\centering
\begin{tabular}{l l ccc}
\hline
Feature Set & F1 Score & Recall & Precision & Accuracy \\
\hline
  API  & 57\% & 58\% & 57\% & 91\% \\
 OPCODE & 65\% & 64\% & 66\% & 93\% \\
 PERM & 81\% & 84\% & 78\% & 96\% \\
 API+OPCODE & 70\% & 68\% & 72\% & 94\% \\
  API+PERM & 82\% & 82\% & 83\% & 96\% \\
 OPCODE+PERM & 87\% & 85\% & 89\% & 97\% \\
 \textbf{ALL} & \textbf{89\%} & \textbf{89\%} & \textbf{89\%} & \textbf{98\%} \\
ALL (RF) & 83\% & 73\% & 97\% & 97\% \\
\hline
\end{tabular}
\label{table:performance_comparison}
\end{table}

In the following, we focus on the analysis of the results of the LR classifier on the full feature set. To assess the limitations of our approach, we examined false-negative (FN) malware samples among the 400 malicious apps in our test set. Using VirusTotal behavior reports, we extracted the corresponding MITRE ATT\&CK signatures for each sample. MITRE ATT\&CK is an openly accessible framework that organizes adversary tactics and techniques derived from real-world threat activity. Tactics are the reasons why an attacker performs an action whereas techniques describe how adversaries achieve their tactical goals by carrying out specific actions. MITRE provides a dedicated matrix for Android platform that includes the list of tactics and techniques~\cite{mitre}. Our analysis revealed notable behavioral differences between FN samples and true positives (TPs):
\subsubsection{Obfuscation Detection}
Using VirusTotal’s obfuscation tag, we found that approximately 71\% of all malware were labeled as obfuscated; among these, 89.7\% were correctly detected (TPs), while only 10.3\% were missed (FNs). We observe a similar trend using MITRE ATT\&CK mappings (technique T1406: \textit{Obfuscated Files or Information}) extracted from VirusTotal reports, where 68\% of all malware samples exhibited obfuscation-related \textit{defense-evasion} behavior. Again, about 90\% of these samples were correctly identified as TPs. In general, most of the obfuscated malware was successfully detected by the LR classifier, with only a small fraction evading detection. This demonstrates the effectiveness of our BYOL-based pre-training combined with LR classification in handling obfuscated malware.


\subsubsection{Attack Patterns}
Table~\ref{tab:attack_patterns} shows tactic prevalence in FNs and TPs for our Android detector. \textit{Collection} and \textit{Discovery} exhibit uniformly high rates in both FNs and TPs (around 80\%), indicating these behaviors are widespread and thus weakly discriminative. In contrast, \textit{Command and Control} (80.2\% TP vs. 60.9\% FN) and Credential Access (40.1\% TP vs. 8.7\% FN) are more prevalent among TPs, suggesting our proposed model is particularly effective in detecting communication and credential theft signals. 

In terms of specific techniques, we focus on the most significant patterns due to space limitations. Within \textit{command and control}, TP samples demonstrated higher usage of application layer protocols (80\% TP vs. 59\% FN) and encrypted channels (80\% TP vs. 59\% FN). For \textit{collection}, TP samples showed more intrusive behaviors including audio recording (45\% TP vs. 35\% FN), clipboard monitoring (20\% TP vs. 4\% FN), and accessing sensitive device logs (25\% TP vs. 0\% FN). FN samples exhibited notable \textit{defense evasion} through device data deletion (18\% TP vs. 28\% FN), suggesting active evidence removal. Within \textit{discovery}, patterns were mixed: FN samples favored basic system information gathering (68\% TP vs. 80\% FN), while TP samples conducted more comprehensive software discovery (54\% TP vs. 41\% FN). 

Overall, these results motivate augmenting evasion-aware features and integrating dynamic/network telemetry, while treating \textit{Collection/Discovery} as baseline behaviors that require stronger contextual cues to improve separability.

\begin{table}[t]
\caption{Main tactics used by attackers}
\centering
\begin{tabular}{l c c }
\hline
\textbf{Tactic} & \textbf{FN} (\%) & \textbf{TP} (\%)  \\
\hline
Collection & 82.6 & 79.9  \\
Command and Control & 60.9 & 80.2  \\
Credential Access & 8.7 & 40.1  \\
Defense Evasion & 76.1 & 70.3  \\
Discovery & 82.6 & 80.8 \\
Impact & 28.3 & 20.9  \\
Network Effects & 21.7 & 16.9 \\
\hline
\end{tabular}
\label{tab:attack_patterns}
\end{table}

\section{Conclusion and Future Work}
\label{sec:conclusion}
In this paper, we addressed temporal bias in ML-based Android malware detection by constructing a time-stamped dataset of benign and malicious apps. We also proposed an API call–based timestamp verification method to validate app release time estimation. Using this dataset, we applied time-aware cross-validation with a logistic regression classifier and incorporated BYOL self-supervised learning to enhance static feature representations. Our FN/TP analysis, based on MITRE ATT\&CK behavior patterns from VirusTotal, highlighted the effectiveness of our system in detecting obfuscated malware. Future work will focus on refining feature representations (e.g., adding evasion-aware features) as well as changing the representation of the apps (e.g., converting APKs to images), to further analyze and strengthen real-world malware detection performance. Moreover, we plan to conduct a deeper comparison with prior SSL and representation-learning approaches for Android malware detection, and to evaluate a broader range of classifiers for more comprehensive benchmarking.

\bibliographystyle{ieeetr}
\bibliography{references}

\end{document}